\documentclass[onecolumn,prd,floatfix,preprintnumbers,nofootinbib]{revtex4-2}

\usepackage{graphicx} % Required for inserting images
\usepackage{amsmath,amssymb}
\usepackage{siunitx}
\usepackage{hyperref}
\usepackage{braket}
\usepackage{wrapfig}
\usepackage{subcaption}
\newcommand{\dint}{{\rm d}}

\newcommand{\xpom}{x_{I\!\!P}}

\begin{document}

\author{Tobias Toll, Nahid Vasim}
\affiliation{
Department of Physics, Indian Institute of Technology Delhi, Hauz Khas, New Delhi 110 016, India
}

\title{
Determining the Proton's Gluon Cloud Geometry from HERA Data
}

\preprint{}

\begin{abstract}
The transverse shape of the proton's small-$x$ gluon distribution is determined from exclusive $J/\psi$ photoproduction at HERA. We derive analytic expressions for the coherent and incoherent diffractive cross sections in the hotspot model at leading twist, enabling a global fit to all 104 available H1 and ZEUS data points, with $|t|<30~{\rm GeV}^2$ and $40\leq W\leq 251~{\rm GeV}$, with $\chi^2/{\rm ndf}=0.72$. The gluonic hotspots are resolved into a core of size 0.088(10) fm surrounded by an exponential halo of range 0.20(2) fm, matching the penetration depth of lattice flux tubes. This shape is stable over $\xpom$ and over the assumed number of hotspots in the range $5\leq N_{hs}\leq 12$. We find the Ginzburg-Landau parameter of the QCD vacuum to be $\kappa=2.7(7)$, which corresponds to fluxtube geometry in a type-II dual superconductor. The calculated QCD string-tension from the extracted parameter values is in agreement with theoretical calculations on the lattice. The gluonic geometry evolves only through a slow transverse diffusion of the hotspot centers with $\alpha'_{\rm eff.}=0.060(19)~{\rm GeV}^{-2}$, while the incoherent cross section at small $|t|$ is dominated by shot-noise fluctuations at the order of a few gluons per hotspot.

\vspace{1ex}
\begin{center}
\fbox{\parbox{0.85\linewidth}{\small
\textbf{Note} This preprint is superseded by T.~Toll and N.~Vasim, \emph{Measuring Dual-Superconductor Length Scales in the QCD Vacuum at HERA}, arXiv:2609.28700, and will not be submitted for publication. Known issues: the quoted fit uncertainties are too small by a
factor of about 1.4, because of an error in the normalization of the Hessian, and the dipole-model parameters were not consistently used. The analytic results of Sec.~\ref{sec:calc} are unaffected and are perhaps more pedagogically reported here. 
The central values change little and the main conclusions hold, with three exceptions: the saturation-scale fluctuations correspond to about five gluons in the proton, i.e. fewer than one per hotspot at $N_{hs}=8$, whether $\lambda$ sets the scale of $\alpha_S$ depends on the value assumed for the string tension; and the data favour the Gaussian core by $\Delta\chi^2\approx2.6$, not 5. Please take numerical results from \cite{Toll:2026zuh}}}
\end{center}
\end{abstract}

\maketitle

\section{Introduction}
The geometric structure of the hadronic matter is a fundamental question in QCD. At large energy experiments, where the smallest momentum fractions of the protons are probed, this becomes the question of the spatial structure of the proton's gluon cloud, resulting from the strong force binding its valence structure in a coherent hadron. While this question is interesting on its own, its answer also affects other physics, such as QCD flow dynamics of the quark-gluon plasma in heavy ion collisions at RHIC and the LHC, where the proton structure serves as an initial state of subsequent dynamics~\cite{Schenke:2012hg,Schenke:2012wb,Mantysaari:2017cni}. This is also expected to be a key measurement at the Electron-Ion Collider (EIC)\cite{Accardi:2012qut,AbdulKhalek:2021gbh}.

Currently, the most fertile ground to answer this question comes from the H1 and ZEUS experiments at HERA, where electrons and protons collided at center of mass energies up to $\sqrt{s}=318~\rm{GeV}$. Especially exclusive diffractive events are sensitive to the geometrical structure of the proton, since these events probe the Mandelstam $t$ variable, and $\Delta=\sqrt{-t}$ is the Fourier conjugate of the spatial structure of the proton. 

For large values of $|t|$, the cross section is particularly sensitive to  event-by-event geometrical fluctuations in the initial state of the proton. In these \emph{incoherent} events, the proton dissociates. For small values of $|t|$, however, the cross-section is dominated by \emph{coherent} events, described by the square of the first moment of the amplitude, in which the proton stays intact. The incoherent cross section has been reasonably well described using a constituent quark model where the gluon cloud is concentrated in three hotspots surrounding the valence quarks acting as sources for small-$x$ gluons (where $x$ is the gluons' fraction of the proton's momentum)~\cite{Mantysaari:2016ykx,Mantysaari:2016jaz,Cepila:2016uku}. In previous studies, the hotspot profile has been assumed to be Gaussian. These models have had two shortcomings: they fail to describe the incoherent spectrum for $|t|>2$~GeV$^2$ on their own, and they fail to simultaneously describe both the coherent and incoherent $t$-spectra. To address the latter, a Bose-Einstein like correction was introduced \cite{Kumar:2021zbn} which made the hotspot shape more peaked in the center. To address the former, in \cite{Kumar:2024kns} a hotspot evolution was introduced which gave a $1/|t|$ power spectrum which could be fitted to the incoherent data. While only one parameter sufficed to fit the entire $t$-spectrum, this approach was not well justified by perturbative QCD. In \cite{Demirci:2022wuy}, the McLerran-Venugopalan model \cite{McLerran:1993ni,McLerran:1993ka} was used, where pointlike fluctuating sources are placed inside the Gaussian hotspots,  and they could thus get a reasonable qualitative description of the entire incoherent $t$ spectrum. However, a consistent precise description of all HERA $J/\psi$ data could not be found. Later studies, evolving the target using the JIMWLK evolution~\cite{Jalilian-Marian:1997gr,JalilianMarian:1997dw,Weigert:2000gi,Iancu:2000hn,Iancu:2001ad}, could describe dense systems in ultra-peripheral Pb-Pb collisions at the LHC for $|t|<1~{\rm GeV}^2$ \cite{Mantysaari:2025ltq} as well as dilute proton targets up to $|t|=8~{\rm GeV}^2$, where a simultaneous good description of the coherent cross section was found \cite{Le:2025bvq}.

Other studies of QCD as a dual superconducting condensate, especially on the lattice, suggest that the QCD vacuum confines the chromo-electric field into fluxtubes whose transverse profile has an exponential halo corresponding to the field penetration depth $\lambda$ of the condensate, combined with a core of a size $\xi$ corresponding to the correlation length of the condensate \cite{Luscher:1980iy, Shibata:2012ae, Shibata:2019nit, Cea:2012qw, Cea:2014uja}. These are solutions to the Ginzburg-Landau equation, which has as an approximate analytic solution the Clem distribution \cite{Clem:1975ohd}. 
The correlation length corresponds to a scalar particle mass, in the QCD vacuum the natural candidate is a scalar glueball, while the halo corresponds to a vector particle mass, which for $\lambda>\xi/\sqrt{2}$ has a vector gluelump as the natural candidate. Another recent approach to find the QCD flux-tube geometry is from the entanglement entropy \cite{Amorosso:2024jhep, Amorosso:2026mdo}, where the flux-tube is treated as a vibrating string.

In this paper, after an overview of the dipole model in the next section, we will present an analytical calculation of the coherent and incoherent cross sections applicable to dilute systems in section \ref{sec:calc}. Here we detail the possible hotspot profiles and the overall energy dependence of the proton's geometry. In section \ref{sec:HERA} we confront the models to the HERA data, and have a detailed discussion on the physical meaning of all the extracted parameters. Section \ref{sec:HERA} also contains several systematic tests of the extracted gluon cloud geometry. We end with a summary and outlook in section \ref{sec:conclusions}.

\section{Exclusive Diffraction in the Dipole Model}
We study the exclusive production of $J/\psi$ vector mesons in the dipole picture~\cite{Nikolaev:1990ja,Mueller:1993rr} in electron-proton photoproduction. The electron and proton interact via the exchange of a quasi-real photon, and a single $J/\psi$ meson is produced. In exclusive diffractive scattering there is no exchange of quantum numbers and the proton stays intact through the interaction. However, in incoherent scattering, the proton gets excited and subsequently dissociates, the remnants of which can be measured by detectors placed near the beampipe.

In the dipole model, the virtual photon fluctuates into a quark antiquark color dipole, which interacts with the proton via exchange of one or many pairs of gluons before it recombines into the vector-meson. 
The total photon-proton cross-section for the interaction can be written as:
\begin{eqnarray}
    \frac{\dint\sigma^{\gamma^*p}}{\dint t}=\frac{1}{16\pi}\braket{|\mathcal{A}^{\gamma^*p}|^2}
\end{eqnarray}
where $\braket{\cdot}$ denotes the averaging over initial state degrees of freedom in the target proton. In photoproduction, the amplitude is~\cite{Kowalski:2006hc}:
\begin{eqnarray}
    \mathcal{A}^{\gamma^*p}(\xpom, \Delta)&=&i\int\dint^2\vec r\int \frac{\dint z}{4\pi}\int\dint^2\vec b e^{-i(\frac12-z)\vec r\cdot\vec\Delta}e^{-i\vec\Delta\vec b}(\Psi^*\Psi_V)(\vec r, z)
     \mathcal{N}_{\rm dip.}(\vec b, \vec r, \xpom)
\end{eqnarray}
Here, $\vec r$ is the 2-vector between the quark and antiquark in the dipole, $\vec b$ is the impact parameter between the dipole and the proton, $z$ is the fraction of the photon's lightcone momentum taken by the quark. The exchanged gluons' momentum fraction $\xpom=(M_V^2-t)/W^2$, where $W$ is the photon-proton invariant mass, and $M_V$ is the vector meson mass. The exchanged momentum is $\vec\Delta$, where the Mandelstam variable $t=-\vec\Delta^2$. Here, $(\Psi^*\Psi_V)$ denotes the wave-overlap between the initial state photon and the final state vector meson. We use the Boosted Gaussian wave function~\cite{Nemchik:1994fp,Kowalski:2006hc} for the vector meson. The amplitude constitutes a Fourier transform from $\vec{b}$ to $\vec\Delta$, which means that the differential cross section acts as a probe for spatial structure in the proton.
In the Good-Walker picture~\cite{Good:1960ba}, the coherent cross-section is given by the first moment of the amplitude, while the incoherent cross-section can be written as the variance of the amplitude:
\begin{eqnarray}
        \frac{\dint\sigma^{\gamma^*p}_{\rm coh.}}{\dint t}=\frac{1}{16\pi}\big|\braket{\mathcal{A}^{\gamma^*p}}\big|^2,~~
        \frac{\dint\sigma^{\gamma^*p}_{\rm inc.}}{\dint t}=\frac{1}{16\pi}\left(\braket{|\mathcal{A}^{\gamma^*p}|^2}-\big|\braket{\mathcal{A}^{\gamma^*p}}\big|^2\right)
\end{eqnarray}
In the IPsat model~\cite{Kowalski:2003hm,Rezaeian:2012ji, Rezaeian:2013tka, Mantysaari:2018nng, Sambasivam:2019gdd}, the dipole amplitude $\mathcal{N}_{\rm dip.}$ is written:
\begin{eqnarray}
    \mathcal{N}_{\rm dip.}^{\rm sat}=1-\exp\left(-\frac{r^2\pi^2}{2N_C}\alpha_S(\mu^2)\xpom g(\xpom,\mu^2)T_p(\vec b)\right)
\end{eqnarray}
where $\alpha_S$ is the strong coupling, $g$ the DGLAP evolved gluon density at the scale $\mu^2=C/r^2+\mu_0^2$. The proton's thickness function $T_p$ is the main focus of this paper. This amplitude saturates at large gluon densities, large dipole sizes, and at a large proton thickness. We use the parameter values from \cite{Mantysaari:2018nng}.

It has been shown that for HERA measurements, both inclusive and exclusive~\cite{Mantysaari:2018nng,Lappi:2010dd,Sambasivam:2019gdd, Kumar:2024kns}, the leading twist expansion of the IPsat model can describe the data well, indicating a small level of saturation. The IPnonsat model is given by:
\begin{eqnarray}
    \mathcal{N}_{\rm dip.}^{\rm nosat}=\frac{r^2\pi^2}{2N_C}\alpha_S(\mu^2)\xpom g(\xpom,\mu^2)T_p(\vec b)
\end{eqnarray}

Here, only the imaginary part of the amplitude is included. One can take the real part into account by multiplying the cross section by a factor $(1+\beta^2)$ where $\beta$ is the real-to-imaginary ratio of the amplitude given by $\beta=\tan(\lambda_g\pi/2)$, with $\lambda_g=\dint\log\mathcal{A}^{\gamma^*p}/\dint\log1/\xpom$~\cite{Kowalski:2006hc} at leading twist. 

To take into account that the two gluons may carry different momentum fractions $\xpom$ the amplitude is also multiplied by a skewedness correction $R_g$~\cite{Shuvaev:1999ce}, where 
\begin{eqnarray}
    R_g=\frac{2^{2\lambda_g+3}}{\sqrt{\pi}}\frac{\Gamma(\lambda_g+5/2)}{\Gamma(\lambda_g+4)}.
\end{eqnarray}

\section{Analytical Calculation of the Cross Sections}
\label{sec:calc}
In the hotspot model, the proton's gluon content is concentrated into several hotspots~\cite{Mantysaari:2016ykx, Mantysaari:2016jaz}. A physical picture is that they are formed as bremsstrahlung around the three valence quarks, and thus remain geometrically correlated to their sources. They can also form around large $x$ gluons that act as sources for the small $x$ gluons that make up the hotspot. 
Using the independent scattering approximation, 
the proton's thickness function becomes:
\begin{eqnarray}
    T_p(\vec b)=\frac{1}{N_{hs}}\sum_i T_{hs}(\vec b-\vec b_i)
\end{eqnarray}
where the hotspot centers are distributed by $T_c(\vec b)$. All thickness functions ($T_p$, $T_{hs}$, and $T_c$) are normalized to unity. 

In the IPnonsat model, the amplitude is given by
\begin{eqnarray}
    \mathcal{A}^{\gamma^*p}(\xpom, \Delta)&=&\int_0^\infty\dint r\int_0^1\dint zD_{rz}^{\xpom} %\nonumber \\&~&\times
    \int\dint^2\vec b e^{-i\vec\Delta\vec b}T_p(\vec b) \nonumber
\end{eqnarray}
with
\begin{eqnarray}
    D_{rz}(\xpom, \Delta)=(\Psi^*\Psi_V)(r, z)\frac{r^3\pi^2}{4 N_C}\alpha_S(\mu^2)\xpom g(\xpom,\mu^2)J_0\left(\left(\frac12\!-\!z\right) r\Delta\right)\nonumber
\end{eqnarray}
We see that the $\vec b$-dependence factorizes. The Fourier transform over $\vec b$ can be written as a sum over hotspot form factors which are identical:
\begin{eqnarray}
    \tilde T_p(t)=\frac{\hat T_{hs}(t)}{N_{hs}}
    \sum_{i=1}^{N_{hs}}e^{-i\vec\Delta\vec b_i}, ~~~\hat T_{hs}(t)=\int\dint^2\vec be^{-i\vec\Delta\vec b}T_{hs}(\vec b)
\end{eqnarray}

For IPnonsat, in the Good-Walker picture, the observable coherent and incoherent cross sections are given by:
\begin{eqnarray}
    \frac{\dint \sigma_{\rm coh.}}{\dint t}\propto |\braket{\tilde T_p(t)}|^2, ~~
    \frac{\dint \sigma_{\rm inc.}}{\dint t}\propto \braket{|\tilde T_p(t)|^2}-|\braket{\tilde T_p(t)}|^2
\end{eqnarray}
We assume that the hotspot profile is isotropic around its center. Define the single hotspot characteristic function $\phi(t)$ as:
\begin{eqnarray}
    \phi(t)\equiv\braket{e^{-i\vec \Delta\vec b}}=\int\dint^2\vec b ~T_c(\vec b)e^{-i\vec b\vec \Delta}
\end{eqnarray}
If we assume that $T_c$ is Gaussian with width $B_{c}$, then $\phi(t)=\exp(-B_{c}|t|/2)$. Assuming that the hotspot positions $\vec b_i$ are independent and identically distributed, the first moment of the thickness function becomes:
\begin{eqnarray}
    \braket{\tilde T_p(t)}=\frac{\hat T_{hs}(t)}{N_{hs}}
    \sum_{i=1}^{N_{hs}}\braket{e^{-i\vec\Delta\vec b_i}}=\hat T_{hs} e^{-\frac{B_{c}|t|}{2}}
\end{eqnarray}
For the second moment of the thickness function we get:
\begin{eqnarray}
    \braket{|\tilde T_p(t)|^2}&=&\frac{\hat T_{hs}^2}{N_{hs}^2}
    \sum_{i=1}^{N_{hs}}\sum_{j=1}^{N_{hs}}
    \braket{e^{-i\vec\Delta(\vec b_i-\vec b_j)}} %\nonumber \\
    = 
    \frac{\hat T_{hs}^2}{N_{hs}^2} \left[N_{hs}+N_{hs}(N_{hs}-1)\braket{e^{-i\vec\Delta\cdot\vec b_i}}\braket{e^{+i\vec\Delta\cdot\vec b_j}}\right]\nonumber \\
    &=&\frac{\hat T_{hs}^2}{N_{hs}}
    \left[1+(N_{hs}-1)e^{-B_{c}|t|}\right]\equiv \frac{\hat T_{hs}^2}{N_{hs}}
    \left[1+(N_{hs}-1)G(t)\right]
\end{eqnarray}
Where in the last step we define the pair-correlator $G(t)=\braket{\exp(-i\vec\Delta(\vec b_i-\vec b_j))}$ for $i\neq j$. For a Gaussian hotspot distribution it is $G(t)=\exp(-B_c|t|)$, and for independent hotspots $\phi^2=G$. The variance that goes into the incoherent cross section is:
\begin{eqnarray}
    \braket{|\tilde T_p(t)|^2}-\braket{\tilde T_p(t)}^2 = 
    \frac{\hat T_{hs}^2}{N_{hs}}\left(1-e^{-B_{c}|t|}\right)
\end{eqnarray}

This is the variance due to the positions of the hotspots in the initial state. However, this expression does not take into account that the center-of-mass (CM) of the proton should be at the origin, which is explicitly seen by the fact that the variance is non-zero for $N_{hs}=1$.
We can take this into account by shifting the hotspot positions in each configuration such that their center-of-mass is at the origin. It should be noted that the second moment of the thickness function, and the pair-correlator $G$, only depend on the relative positions of the hotspots, and are unaffected by this shift. 

For the first moment, we have:
\begin{eqnarray}
    e^{-i\vec\Delta\cdot\vec b'_i}=e^{-i\frac{N_{hs}-1}{N_{hs}}\vec\Delta\cdot\vec b_i}
    \prod_{k\neq i}e^{i\frac{\vec\Delta\cdot\vec b_k}{N_{hs}}}
\end{eqnarray}
since 
$\vec b'_i=\vec b_i-\braket{\vec b}=\vec b_i-1/N_{hs}\sum_{i=1}^{N_{hs}}\vec b_i = \vec b_i(1-1/N_{hs})-1/N_{hs}\sum_{k\neq i}\vec b_k$. Here, primed vectors are the positions after the shift. 
The characteristic function for the CM shifted average becomes:
\begin{eqnarray}
     F(t) &\equiv& \braket{e^{-i\vec\Delta\cdot\vec b'_i}}=\braket{e^{-i\frac{N_{hs}-1}{N_{hs}}\vec\Delta\cdot\vec b_i}}\prod_{k\neq i}\braket{e^{i\frac{1}{N_{hs}}\vec\Delta\cdot\vec b_k}} =
    \phi\left(\frac{N_{hs}-1}{N_{hs}}\Delta\right)
    \left[\phi\left(\frac{\Delta}{N_{hs}}\right)\right]^{N_{hs}-1} \nonumber \\
    &=&
    \exp\left(-\frac12B_{c}|t|\frac{N_{hs}-1}{N_{hs}}\right)=G(t)^{\frac{N_{hs}-1}{2N_{hs}}}
    \label{eq:average}
\end{eqnarray}
where in the next-to-last step we assume a Gaussian distribution of the hotspots. The first moment of the thickness function then becomes:
\begin{eqnarray}
    \braket{\tilde T_p'(t)}=\hat T_{hs}F(t)
\end{eqnarray}
The CM-corrected variance then is:
\begin{eqnarray}
%    \braket{|\tilde T_p'(t)|^2}-\braket{\tilde T_p'(t)}^2 = 
    {\rm Var}_{\rm geom.}[\tilde T_p]=\frac{\hat T_{hs}^2}{N_{hs}}\left[1+(N_{hs}-1)G(t)\right]-\hat T_{hs}^2F^2(t)\nonumber \\
    \label{eq:var}
\end{eqnarray}
We see that the variance disappears when $N_{hs}=1$, as well as for $|t|\rightarrow 0$. We also see that the large $|t|$ part of the spectrum is directly governed by $\hat T_{hs}^2/N_{hs}$ (assuming the substructure $\hat T_{hs}$ falls off slower than $\exp(-B_c|t|)$). As $N_{hs}\rightarrow\infty$, the profile approaches the round Gaussian profile, as seen by the disappearance of the variance, and that $F(t)$ becomes a single hotspot Gaussian.

Following \cite{Mantysaari:2016jaz,McLerran:2015qxa}, each hotspot is multiplied by a random factor $\zeta_i$ drawn from a log-normal distribution reflecting event-by-event saturation scale fluctuations. We assume that $\braket{\zeta_i}=1$, which gives $\braket{\zeta_i^2}=\exp(4\lambda_g^2\sigma_S^2)$, such that the coherent cross-section is left unchanged. This is slightly different from what M\"antysaari and Schenke (MS) do, in \cite{Mantysaari:2016jaz}, they use a unit median and $\braket{\zeta_i}=\exp(2\lambda_g^2\sigma_S^2)$. The difference is absorbed by the fitted parameter $\sigma_S$. 

The total variance is:
\begin{eqnarray}
    {\rm Var}_{\rm total}&=&{\rm Var}_{\rm geom.}[\tilde T_p]+\rm {Var_{sat.}}[\zeta_i], ~~ {\rm Var_{sat.}}[\zeta_i]=\left(e^{4\lambda_g^2\sigma^2_S}-1\right)\frac{\hat T_{hs}^2}{N_{hs}}
    \label{eq:finalvariance}
\end{eqnarray}
For small $|t|$, the main contribution to the incoherent cross section comes from saturation scale fluctuations, while for large $|t|$, $F, G\rightarrow0$ and the relative contribution from the two sources approach a constant where ${\rm Var}_{\rm sat.}/{\rm Var}_{\rm geom.}=\exp(4\lambda_g^2\sigma_S^2)-1$. 

We checked the analytic results presented in this section with a numerical calculation where we randomly generated hotspot configurations, both for the first and second moments of the thickness function, for different hotspot profiles, and found perfect agreement.

\subsection{The Hotspot Profiles}
\label{sec:profiles}
Taking inspiration from work on flux-tube dynamics  \cite{Luscher:1980iy,  Amorosso:2024jhep, Amorosso:2026mdo}, we model the hotspot thickness as a convolution of a Gaussian core and a Bessel function, the latter being the 2D projection of a 3D Yukawa profile $\exp(-r/\lambda)/r$:
\begin{eqnarray}
    T_{hs}(b)=\int\dint^2\vec b'T_G(\vec b-\vec b')T_{K0}(b')
    \label{eq:conv}
\end{eqnarray}
with
\begin{eqnarray}
    T_G(b)=\frac{1}{2\pi B_{hs}}e^{-\frac{b^2}{2B_{hs}}},~
    T_{K0}(b)=\frac{1}{2\pi\lambda^2}K_0\left(\frac{b}{\lambda}\right)
\end{eqnarray}
Then the hotspot formfactor becomes:
\begin{eqnarray}
    \hat T_{hs}(t)=\frac{e^{-B_{hs}|t|/2}}{1+\lambda^2|t|}
    \label{eq:halo}
\end{eqnarray}
The hotspot centers are distributed by a Gaussian with width $B_{c}$.

We may also directly fit the Clem formula \cite{Clem:1975ohd}:
\begin{eqnarray}
    T_{hs}^{\rm Clem}(\vec b)\propto K_0\left(\frac{\sqrt{b^2+\xi_v^2}}{\lambda}\right), ~~~\hat T_{hs}^{\rm Clem}=\frac{1}{\lambda\sqrt{1/\lambda^2+|t|}}\frac{1}{K_1(\xi_v/\lambda)}K_1\left(\xi_v\sqrt{1/\lambda^2+|t|}\right)
    \label{eq:Clem}
\end{eqnarray}
Similarly to eq.~\eqref{eq:conv}, this distribution has a core given by $\xi_v$ and an exponential halo defined by $\lambda$. The interpretive benefit of Clem's formula is that it describes flux-tubes in a superconductor, in our case, the dual-superconducting vacuum of QCD. We can use the result to find the Ginzburg-Landau parameter $\kappa$ \cite{Clem:1975ohd}:
\begin{eqnarray}
    \kappa=\frac{\sqrt{2}}{\alpha}\sqrt{1-\left(\frac{K_0(\alpha)}{K_1(\alpha)}\right)^2}, ~~\alpha=\frac{\xi_v}{\lambda}
    \label{eq:GL}
\end{eqnarray}
where $\kappa<1/\sqrt2$ for a Type-I superconductor, and $\kappa>1/\sqrt2$ for a Type-II superconductor. Further, the Clem parameters can be interpreted in terms of vacuum excitations, where $\sim 1/\xi$ corresponds to a scalar excitation, for which in QCD the natural candidate is a scalar glueball, and $\sim1/\lambda$ corresponds to a lighter vector excitation, for which the natural candidate is a vector gluelump.

While the hotspot profiles described above in eqs.~\eqref{eq:conv}- \eqref{eq:Clem} are our default in this paper, we also considered alternative profiles. As mentioned, the usual profile for hotspots in the literature has always been Gaussian with width $B_{hs}$. We also consider some alternatives in our fits, in order to quantify how sensitive the HERA data are to the exact shape of the hotspot tail. Therefore we convolute the Gaussian core with the following three alternative tails:
\paragraph{Exponential or Dipole tails:}
We let $\hat T_{hs}$ be:
\begin{eqnarray}
    \hat T_{hs}(t)=\frac{e^{-B_{hs}|t|/2}}{\left(1+\lambda^2|t|\right)^n}
    \label{eq:shape2}
\end{eqnarray}
where $B_{hs}$, $\lambda$, and $n$ are free parameters. If $n=1$ we retain the exponential halo profile in eq.~\eqref{eq:halo}, if $n=2$ it would correspond to the dipole formfactor.
\paragraph{Stretched Exponential Tails:}
Here we instead let $\hat T_{hs}$ take the stretched exponential form:
\begin{eqnarray}
        \hat T_{hs}(t)=e^{-B_{hs}|t|/2}e^{-\left(\lambda^2|t|\right)^\beta}
        \label{eq:shape3}
\end{eqnarray}
treating $B_{hs}$, $\lambda$, and $\beta$ as free parameters. If $\beta=0.5$ this would correspond to a power law tail of the hotspot with $T(b)\propto (1+b^2/a^2)^{-3/2}$, while $\beta=1$ corresponds to a Gaussian.
\paragraph{Power law tails:}
The third option is to directly fit a power law tail:
\begin{eqnarray}
    T(b)=\frac{p-1}{\pi a^2}\frac{1}{\left(1+\frac{b^2}{a^2}\right)^p}
\end{eqnarray}
with the Fourier transform:
\begin{eqnarray}
    \hat T_{hs}(t)=e^{-B_{hs}|t|/2}\times \frac{2^{2-p}}{\Gamma(p-1)}\left(a\sqrt{|t|}\right)^{p-1}
    K_{p-1}\left(a\sqrt{|t|}\right)
            \label{eq:shape4}
\end{eqnarray}
with $B_{hs}$, $p$, and $a$ as free parameters. For large $|t|$ this corresponds to an exponentially falling $t$-spectrum. It should be noted that $\braket{b^2}$ is divergent for $p\leq2$ and $\braket{b^2}\sim ~a^2/(p-2)$ for $p>2$, which makes this a separate test than the stretched exponential with $\beta=1/2$.

\subsection{Energy Dependencies}
The HERA data is taken in a large range in $W$ and in order to describe it well, the geometry needs to have an energy dependence. 
We have implemented this in the form of a squeezed proton where we let the proton's width vary as functions of $\xpom$ as 
\begin{eqnarray}
    B_{c}=B_{c0}+2\alpha'\log\left(x_0/\xpom\right)
\end{eqnarray} 
We take $x_0=\xpom(W=90, t=0)=0.0012$, and treat $\alpha'$ as a free parameter. Looking at eq.~\eqref{eq:average}, we see that the effective squeeze becomes $\alpha'_{\rm eff.}=\alpha'(N_{hs}-1)/N_{hs}$. Running coupling BFKL calculations predict that $\alpha'_{\rm eff.}$ is small and positive \cite{Nikolaev:1994kw}. Lotter \cite{Lotter:1996vk} predicts $\alpha'_{\rm eff.}\simeq 0.05~\rm{GeV}^{-2}$. This is smaller than the values reported from ZEUS ($\alpha'=0.115\pm 0.018^{+0.008}_{-0.015}$) \cite{Chekanov:2002xi} and H1 ($\alpha'=0.164 \pm 0.028 \pm 0.030$)\cite{Aktas:2005xu}.

On top of this we also implemented two different models for energy-dependence of the hotspot substructure:
\paragraph{Squeezed Hotspots:}
Similar to previous studies, e.g. \cite{Salazar:2021mpv, Kumar:2022aly}, we may in addition to squeezing the proton, also squeeze the hotspots, such that
\begin{eqnarray}
    B_{hs}(\xpom)=B_{hs0}+2\beta'\log\left(x_0/\xpom\right)
\end{eqnarray}
with $\beta'$ as additional free parameter. Then $\alpha'_{\rm eff.}=(N_{hs}-1)/N_{hs}\alpha'+\beta'$.

\paragraph{Varying Hotspot Number Model:}
Taking inspiration from \cite{Cepila:2016uku}, we alternatively let the average number of hotspots vary with $\xpom$, as $\braket{N_{hs}}=N_{hs0}(x_0/\xpom)^{\delta_N}$, where $\braket{N_{hs}}$ replaces $N_{hs}$ in the cross section calculations. Here $\delta_N$ is treated as a free parameter. This is combined with the overall squeezed proton but separate from the squeezed hotspots. In this model $\alpha'_{\rm eff.}=(\braket{N_{hs}}-1)/\braket{N_{hs}}\alpha'+B_{c}\delta_N/(2\braket{N_{hs}})$. To fully implement this model we would need to modify eq.~\eqref{eq:finalvariance} by adding another term corresponding to the variance of $N_{hs}$. This effect is not included in this paper, as most of the energy dependence is found in the coherent data, and the fit results shown in section \ref{sec:energy} do not warrant a more detailed implementation.

\section{Comparisons to HERA data}
\label{sec:HERA}
\begin{figure}
    \centering
        \includegraphics[width=1\linewidth]{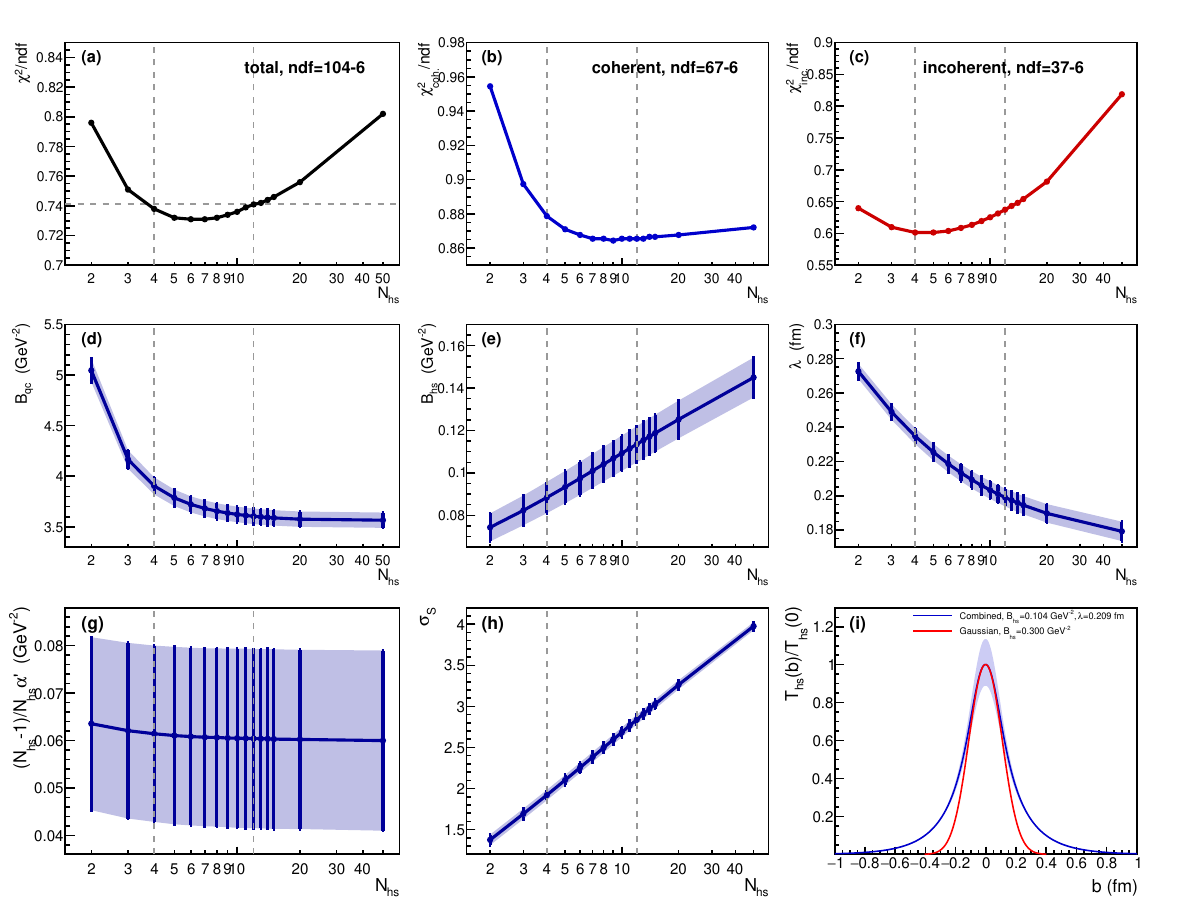}
    \caption{The fit results for the Gaussian core convoluted with a $K_0$ tail, as a function of $N_{hs}$. (a) Total $\chi^2/{\rm ndf.}$, (b) the coherent $\chi^2_{\rm coh.}/{\rm ndf}$ component of the total, (c) the incoherent $\chi^2_{\rm inc.}/{\rm ndf}$ component of the total, (d) the hotspot center distribution width $B_c$, (e) the hotspot width $B_{hs}$, (f) exponential tail $\lambda$, (g) $\alpha'_{\rm eff.}$, (h) the size of the saturation scale fluctuations $\sigma_S$, and (i) the resulting hotspot profile compared with a Gaussian.}
    \label{fig:results_6params} 
\end{figure}
We show the results from the convoluted Gaussian core with exponential tails in fig. \ref{fig:results_6params}, depicting the fit results of the five parameters $B_{c0}$, $\alpha'_{\rm eff.}$, $B_{hs}$, $\lambda$, and $\sigma_S$ scanned over the number of hotspots $N_{hs}$. The uncertainties are obtained by varying the parameters such that the variation $\Delta\chi^2=1$. We fit to exclusive photoproduction $J/\psi$ data from the H1 \cite{Aktas:2005xu,Alexa:2013xxa,Aktas:2003zi} and ZEUS \cite{Chekanov:2002rm,Chekanov:2009ab} experiments at HERA. In total, there are 104 data points, out of which 37 are for incoherent diffraction and 67 are coherent. The center of mass energy and $t$ vary in the ranges $40\leq W\leq 251~{\rm GeV}$ and $0<|t|\leq 30~{\rm GeV}^2$ respectively. This gives a good lever arm in $W$ for the $\alpha'$ parameter as well as in $t$ for the hotspot shape parameters. We also fit the over-all normalization which gives a factor $\sim1.2$, which absorbs well-known sensitivities in the dipole normalization to the charm mass and the vector-meson wavefunction, as well as the normalization uncertainties in the experiments' data sets present in their systematic uncertainties. As this is an overall factor it does not affect the fitted $t$-shapes, and therefore none of the geometric or fluctuation parameters extracted here. 

In fig.\ref{fig:results_6params}(a)-(c) we show the quality of the fit as the reduced chi-squared values for the total fit, as well as the total fit broken up into its coherent and incoherent cross section descriptions respectively. We see that all the data is very well described for all $N_{hs}$, with the best fit at $N_{hs}=6\dots 7$, the best fit range is $N_{hs}=4\dots 12$ being within one standard deviation. The best fit for the coherent data occurs for $N_{hs}=9$, but this distribution is rather flat and has $N_{hs}=4\dots 50+$ within one standard deviation, and as noted above, $N_{hs}\rightarrow\infty$ corresponds to a round proton which is known to describe coherent data well. The incoherent data, which is more sensitive to $N_{hs}$, prefers $N_{hs}=4\dots 5$.

The $\sigma_S$ parameter climbs logarithmically from $1.92(6)$ to $2.84(6)$ over the best fit range. To compare with the results from the MS model \cite{Mantysaari:2016ykx}, which gives $\sigma_{S}^{\rm MS}=0.5$, we need to take into account that our definitions are slightly different, which gives $\sigma_S\approx \sigma_S^{\rm MS}/(2\lambda_g)\sim 2.8\sigma_S^{\rm MS}$ with $\lambda_g=0.18$. At 3 hotspots, we have $\sigma_S=1.7$. If we do not perform the center-of-mass shift, we get $\sigma_{S}=1.6$, which corresponds to $\sigma_S^{\rm MS}=0.6$, which is consistent with their reported value within uncertainties. 

Our value $\sigma_S=2.5$ for $N_{hs}=8$ corresponds to ${\rm Var}[\zeta_i]/\braket{\zeta_i}^2=\exp(4\lambda_g^2\sigma_S^2)-1\approx 0.53\dots 1.72$, with $\lambda_g=0.13\dots0.20$. For a Poisson distribution this would be the inverse of the average occupancy in each hotspot, which corresponds to $1/{\rm Var}[\zeta]=0.6\dots1.9$ gluons per hotspot. However, $\zeta_i$ are generated from a lognormal distribution, which has longer tails, which means that these numbers would be slightly higher. From another perspective we can make an order-of-magnitude estimate of the number of gluons per hotspot in the IPnonsat model from the gluon density $N_g\sim\xpom g(\xpom, \mu^2){\rm d}\log(1/\xpom)\sim 6\cdot3.5\sim 20$ at these kinematics, which for 8 hotspots gives $\sim 2.5$ gluons per hotspot. We can interpret such small occupancies as the small $|t|$ variance being dominated by shot-noise in the hotspot occupancy. From this analogy, the logarithmic growth of $\sigma_S$ as a function of $N_{hs}$ can be directly understood physically.

The hotspot geometry stays remarkably stable throughout the best fit interval, with $B_{hs}=0.088(7)\dots0.114(9)~{\rm GeV}^{-2}$, and $\lambda=0.199(5)\dots0.235(5)~{\rm fm}$. 
We show the resulting profile in fig.\ref{fig:results_6params}(i), with uncertainty bands, and compare it to the MS Gaussian hotspot with width $0.3~{\rm GeV}^{-2}$. We will discuss these values further in the next section.

Within the best fit range, the hotspot centers are distributed with a width $B_{c0}=3.60(8)\dots3.91(8)~{\rm GeV}^{-2}$. We can use the fit results to calculate the gluon size of the proton from:
\begin{eqnarray}
    \braket{b^2}=-4\frac{\dint\tilde T_p(t)}{\dint |t|}\bigg|_{t=0}= 2B_c\frac{N_{hs}-1}{N_{hs}}+2B_{hs}+4\lambda^2
    = \bigg[0.658(4)_{\rm fit}(12)_{N_{hs}}~{\rm fm}\bigg]^2
    \label{eq:GKsize}
\end{eqnarray}
at $\xpom=x_0$, which can be compared with the proton's transverse charge radius 0.6261(6) fm obtained from the Dirac form factor \cite{Burkardt:2002hr, Miller:2007uy} from the 3D charge radius $r_p=0.84075(64)$ fm~\cite{Mohr:2024kco}, where  
\begin{eqnarray}
    \braket{b_p^2} = -4\frac{dF_1}{dQ^2}\bigg|_{Q^2=0} = \frac{2}{3}r_p^2 - \frac{F_{2,p}(0)}{m_p^2}
\end{eqnarray}

\begin{figure}
    \centering
    \includegraphics[width=1\linewidth]{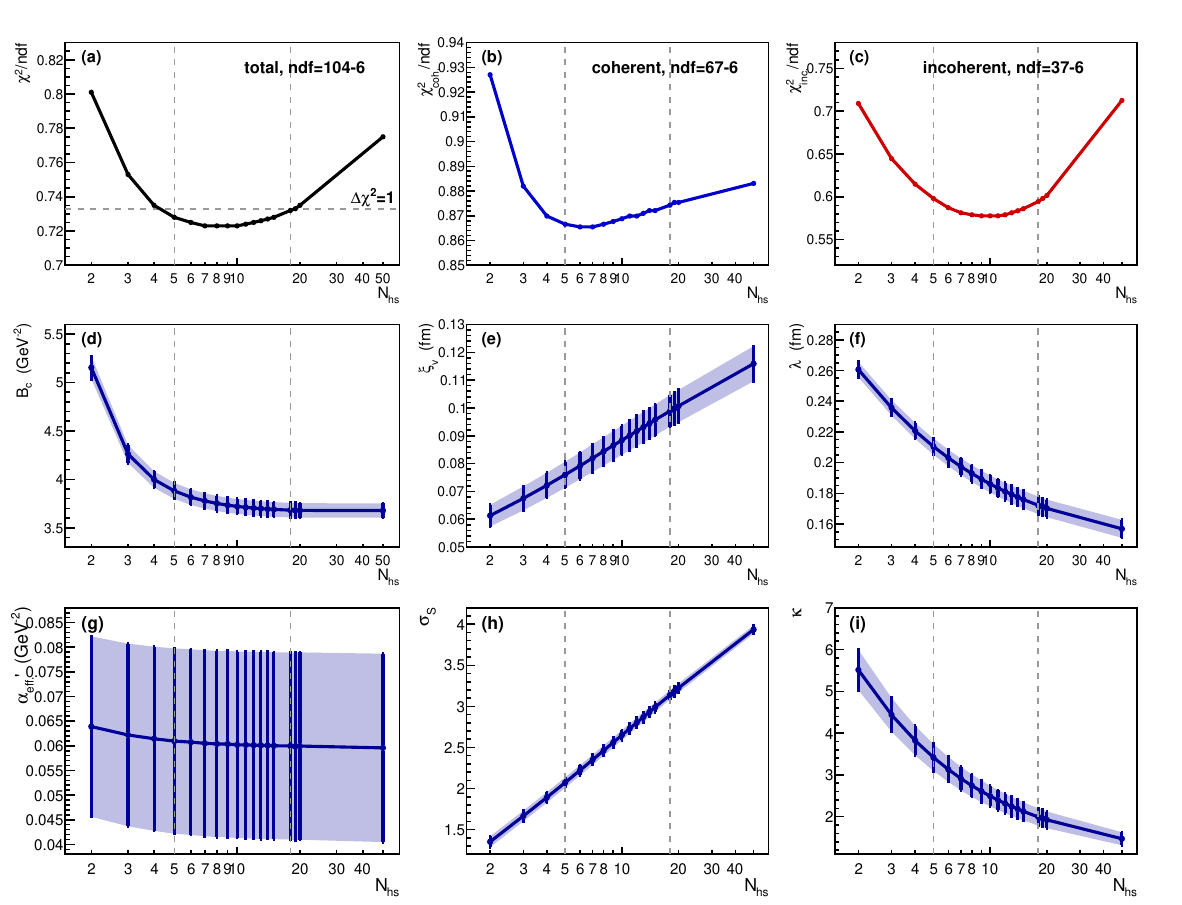}
\caption{Fit results for the Clem profile as a function of $N_{hs}$. (a) Total $\chi^2/{\rm ndf.}$, (b) the coherent $\chi^2_{\rm coh.}/{\rm ndf}$ component of the total, (c) the incoherent $\chi^2_{\rm inc.}/{\rm ndf}$ component of the total, (d) the hotspot center distribution width $B_c$, (e) the core size parameter $\xi_v$, (f) the penetration depth $\lambda$, (g) $\alpha'_{\rm eff.}$, (h) the size of the saturation scale fluctuations $\sigma_S$, and (i) the resulting GL parameter $\kappa$.}
    \label{fig:ClemResults}
\end{figure}
\subsection{The Clem Distribution}
\label{sec:Clemres}
We fitted the Clem distribution in eqs.~\eqref{eq:Clem}. The results are shown in Fig.~\ref{fig:ClemResults}. Here we see that the overall results are very similar to the results from the Gaussian core. The 1$\sigma$ range is in the interval $5\leq N_{hs}\leq 18$ which is larger than in the previous case. Here, the coherent cross section prefers $6\dots 7$ hotspots while the incoherent prefers $10$, it should be noted that the $\chi^2$ curves are rather shallow, and allow for large ranges in $N_{hs}$. The Clem distribution gives a similar fit quality compared to the Gaussian core fit, improvement is only $\Delta\chi^2=0.83$ at $N_{hs}$ values of 8 and 7 for the Clem distribution and the Gaussian-$K_0$ convolution respectively.

\begin{figure}  
    \centering
    \includegraphics[width=1\linewidth]{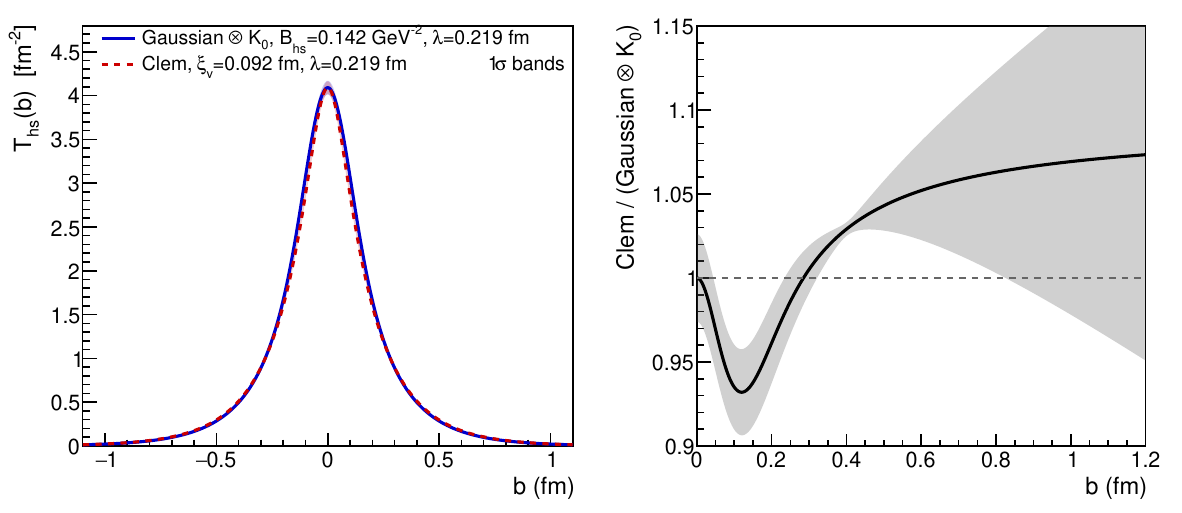}
    \caption{Comparison between the Gaussian and $K_0$ convoluted profile and the Clem profile.}
    \label{fig:comparison}
\end{figure}
In fig.~\ref{fig:comparison} we compare the resulting profiles using the weighted average in the best fit region for the Gauss-$K_0$ convolution, and the parameters that match the Clem distribution at $b=0$ and the same $\lambda$ values. We see that the two profiles are very similar. It should be noted that for all fitted values at different $N_{hs}$, $\xi_v=\sqrt{2B_{hs}}$ within uncertainties,  which corresponds to the size of the core, which means that this number is more important for the fit than the hotspot size at $b=0$. Over the best fit regions, $\sqrt{2B_{hs}}/\xi_v=1.07(9)$.

We use the fitted values to calculate the Ginzburg-Landau parameter using eq.~\eqref{eq:GL} as seen in fig.~\ref{fig:ClemResults}(i), and we get the value $\kappa=2.0(2)\dots 3.4(3)$ on the best fit interval which, in the picture of a flux tube, would put the hotspots inside the Type-II dual superconductor region. We can then use the usual definition of the Ginzburg-Landau parameter $\kappa=\lambda/\xi$ to find $\xi=\lambda/\kappa=0.062(6)\dots0.086(9)~{\rm fm}$, where $\xi$ now has the interpretation of the correlation length in the QCD condensate, while $\lambda$ is the penetration depth. Following \cite{Nishino:2019bzb}, these numbers have a clear interpretation with $\lambda=1/m_V$, and $\xi=\sqrt{2}/m_S$, where $m_V$ and $m_S$ are vector and scalar particle masses respectively. In our case, over the best fit range, $m_V=1.03(11)~{\rm GeV}$, and $m_S=3.9(7)~{\rm GeV}$. Here, the natural corresponding particle candidates would be a vector gluelump and a scalar glueball. The extracted gluelump mass is consistent with literature, in \cite{Bali:2003jq} the authors calculate the lightest gluelump mass to be $m_{1+-}=0.87(15)~{\rm GeV}$ (one should note that this calculation is scheme dependent, since the static adjoint source carries a linearly divergent self-energy, and the quoted value is in the renormalon-subtracted scheme.). The extracted scalar mass is a factor 2.4(4) larger than what has been calculated on the lattice; Athenodorou and Teper report the lightest glueball mass in SU(3), 3+1 dimensions as $m_{0++}=3.405(21)\sqrt{\sigma_0}=1.651(23)~{\rm GeV}$ \cite{Athenodorou:2020ani}, if we assume a string-tension of $\sqrt{\sigma_{0}}=0.485(6)~{\rm GeV}$ (see below). It should be noted that the identification of mass scales depends on specific assumptions on the functional form of the QCD vacuum potential, and the dual superconductor picture does not rely on these assumptions. 

We note that lattice fits of the Clem profile by Cea et al. \cite{Cea:2012qw, Cea:2014uja} get a comparable penetration depth $\lambda=0.175(6)$~fm (using $\sqrt\sigma=0.420~{\rm GeV}$) but a larger core size $\xi_v\approx0.6$~fm (inferred from their $\kappa$ and $\lambda$ values using eq.~\eqref{eq:GL}) and $\kappa\approx 0.2$, which corresponds to a type-I vacuum. Shibata et al. get $\kappa\approx 0.45\dots 0.48$ \cite{Shibata:2012ae} which is also in the type-I regime. However, they are fitting to hadron scale strings which are longer and vibrate, which makes the string thickness grow logarithmically with the length of the string \cite{Luscher:1980iy,Amorosso:2024jhep}. This effect is partly absorbed by the core-size. Such vibrations are not present in the vacuum response to point sources. 

As $\xi_v$ is so closely associated with the core-size $\sqrt{2B_{hs}}$, we may find the corresponding mass $m_{\rm core}=1/\xi_v=2.3(3)~{\rm GeV}$, or $m_{\rm core}=1/\sqrt{2B_{hs}}=2.2(3)~{\rm GeV}$ in respective best-fit regions, which are a factor of 1.4(2) and 1.3(2) larger than the glueball mass respectively. 

As an independent cross check we can also calculate the string tension using Clem's calculation for the free energy density along a string in a regular superconductor \cite{Clem:1975ohd}:
\begin{eqnarray}
    \epsilon&=&\frac{\Phi}{4\pi}\sqrt2H_cH_{c1}'=\left(\frac{\Phi}{4\pi\lambda}\right)^22\kappa H_{c1},~~~\alpha=\frac{\xi_v}{\lambda} \nonumber \\
\end{eqnarray}
The line-energy $L(\alpha)=2\kappa H_{c1}$ has three contributions: gradient-, condensation-, and field- energies: 
\begin{eqnarray}
        L(\alpha) = 2\kappa H'_{c1}= \frac14+\frac14\kappa^2\alpha^2 + \frac{K_0(\alpha)}{\alpha K_1(\alpha)}= \frac34-\frac12\left(\frac{K_0(\alpha)}{K_1(\alpha)}\right)^2+\frac{K_0(\alpha)}{\alpha K_1(\alpha)}
\end{eqnarray}
where we plugged in the variational result for $\kappa$ in eq.~\eqref{eq:GL} in the last step. 

Following \cite{Koma:2002cv}, in the dual theory the tube carries the chromoelectric flux of a quark with the quark charge $e\vec\omega_k$ and the Dirac condition of $\Phi^2=4\pi Ce^2$ (the factor $4\pi$ comes from converting from Gaussian to natural units), with $e^2=4\pi\alpha_S$.
This gives:
\begin{eqnarray}
\sigma=\frac{C\alpha_SL(\alpha)}{\lambda^2}
\label{eq:sigma}
\end{eqnarray}
Here $C$ is a color factor which we infer from Koma to be $C=(N-1)/(2N)$ for $SU(N)$ which gives $C=1/3$ \cite{Koma:2002cv}. In the best fit $N_{hs}$ interval, we get $L(\alpha)=1.79(19)$. Athenodorou and Teper find $r_0\sqrt{\sigma_{0}}=1.160(6)$\cite{Athenodorou:2020ani} and Sommer gives $r_0=0.472(4)~{\rm fm}$ \cite{Sommer:2014mea}, giving $\sqrt{\sigma_{0}}=0.485(6)~{\rm GeV}$. We may use this to match the theoretical value with $\alpha_S=0.37(9)$. In the leading order running coupling in the $\overline{\rm MS}$ scheme that we use in our dipole model $\alpha_S(1/\lambda^2)=0.36(2)$, and $\alpha_S(1/\xi^2)=0.24(2)$. While these numbers are scheme and order dependent, they indicate that $\lambda$ is the relevant scale for the coupling of the flux tube. 

We obtain $B_{c0}=3.78(10)~{\rm GeV}^{-2}$. We can calculate the gluon size of the proton from the Clem profile by:
\begin{eqnarray}
    \braket{b^2}=2B_c\frac{N_{hs}-1}{N_{hs}}+\lambda^2\left(4+2\alpha\frac{K_0(\alpha)}{K_1(\alpha)}\right)
    \label{eq:clemsize}
\end{eqnarray}
which gives $b_{\rm rms}=0.649(4)_{\rm fit}(10)_{N_{hs}}~{\rm fm}$, which is consistent with the size found for the Gaussian core with an exponential halo above. As $\xi_v, B_{hs}\rightarrow 0$ expressions in eq.~\eqref{eq:clemsize} agree with eq.~\eqref{eq:GKsize}.

\begin{figure}
    \centering
    \includegraphics[width=1\linewidth]{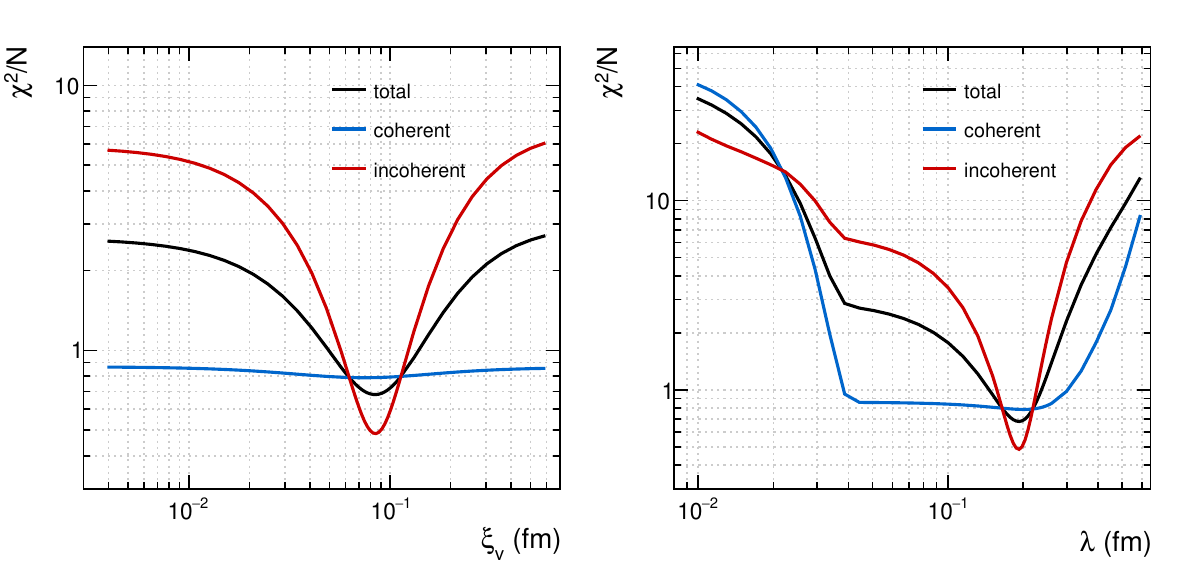}
    \caption{The refitted 5-parameter qualities as a function of fixed $\xi_v$ (left) and fixed $\lambda$ (right). The fit-quality is expressed in $\chi^2/N$ where $N$ is the number of experimental data points (Total=104, Coherent: 67, Incoherent: 37) for $N_{hs}=8$.}
    \label{fig:xilamscan}
\end{figure}
As a further systematic check, we performed a scan of fixed values of $\xi_v$ and $\lambda$ respectively and fitted the remaining 5 parameters for each value, to see how well the data constrain the hotspot geometry, at the best fit $N_{hs}=8$. As expected, the incoherent cross section is very sensitive to these variables, and its fit deteriorates quickly away from the best fitted values, while the coherent data show next-to-no dependency on the hotspot geometry over a large range in both variables. For the core size the coherent data is agnostic throughout the scanned interval. With 95\% confidence level from this scan, the incoherent cross section gives $\xi_v=0.085^{+0.015}_{-0.014}~{\rm fm}$ and $\lambda=0.193^{+0.014}_{-0.016}~{\rm fm}$.

\subsection{Energy Dependence}
\label{sec:energy}
From eq.\eqref{eq:average} we see that the effective $W$-slope parameter $\alpha'_{\rm eff.}=(N_{hs}-1)/N_{hs}\alpha'=0.060(19)_{\rm fit}(1)_{N_{hs}}~{\rm GeV}^{-2}$, which remains constant throughout the interval for both fits, as seen in Figs.\ref{fig:results_6params}(g) and \ref{fig:ClemResults}(g). This value is consistent with the running coupling BFKL predictions of $\alpha'_{\rm eff.}\simeq 0.05~\rm{GeV}^{-2}$\cite{Lotter:1996vk} . We see in table \ref{tab:correlation_matrix1} that $\alpha'$ is completely uncorrelated from the other fit parameters with an absolute correlation $<0.05$, which means that the extracted value is independent of the geometry. It is smaller than the values reported from ZEUS ($\alpha'=0.115\pm 0.018^{+0.008}_{-0.015}~{\rm GeV}^{-2}$) \cite{Chekanov:2002xi} and H1 ($\alpha'=0.164 \pm 0.028 \pm 0.030~{\rm GeV}^{-2}$)\cite{Aktas:2005xu}. We note that the HERA values are extracted from Regge fits, exponential slopes with independent normalization to each bin, while our fit describes the entire spectrum with the normalization and its $W$-dependence, which is connected to the DGLAP-evolved dipole amplitude and the combined core and exponential shape.
\begin{table}
    \begin{center}
        \begin{tabular}{|c|c|c|c|c|}
        \hline
         & Fit 1 & Fit 2 & Fit 3 & Fit 4 \\
                         &  $G\otimes K_0$     & Clem &     $B_{hs}(\xpom)$ & $N_{hs}(\xpom)$  \\
        \hline
        $B_{c0}$ (GeV$^{-2}$)      & 3.66(8)   & 3.75(8)    & 3.70(8)   & 3.66(7)\\
        $B_{hs0}$ (GeV$^{-2}$)     & 0.104(8)  &     ---    & 0.133(22) & 0.105(8)\\
        $\sqrt{2B_{hs}}$ or $\xi_v$ (fm) & 0.090(4) & 0.084(5) & 0.102(8) & 0.090(3) \\
        $\lambda$ (fm)             & 0.209(5)  & 0.193(6)   & 0.203(6)  & 0.209(5) \\
        $\sigma_S$                 & 2.50(6)   & 2.46(6)    & 2.47(7)   & 2.48(7)\\
        $\alpha'$ (GeV$^{-2}$)     & 0.069(22) & 0.069(22) & 0.051(25)  & 0.059(30)\\
        $\beta'$ (GeV$^{-2}$)      &    ---    &      ---   & 0.016(11) &   ---     \\
        $\delta_N$                 &     ---   &     ---    &    ---    & 0.040(81)\\
        \hline\hline
        $\alpha'_{\rm eff.}~({\rm GeV}^{-2})$ & 0.060(19) & 0.060(19) & 0.061(18) & $0.061(31)-0.062(32)$ \\
        \hline\hline
        $\chi^2/{\rm ndf}$ & 71.8/98 & 70.8/98 & 70.5/97 & 71.7/97 \\
        $\chi_{\rm coh.}^2$ & 52.8 & 52.9 & 52.8 & 52.8\\
        $\chi_{\rm inc.}^2$ & 19.0 & 18.0 & 17.7 & 18.9\\
        \hline
        \end{tabular}
    \end{center}
    \caption{Fit results for $N_{\rm hs}=8$ for four different models. There are 104 data points in total out of which, 67 are coherent and 37 incoherent diffraction. In Fits 1 and 2, 6 parameters are fitted, in Fit 3, $B_{hs}(\xpom)=B_{hs0}+2\beta'\log(x_0/\xpom)$ with $\alpha'_{\rm eff}=(N_{hs}-1)/N_{hs}\alpha'+\beta'$, and in Fit 4, $N_{hs}(\xpom)=N_{hs0}(x_0/\xpom)^{\delta_N}$, with $\alpha'_{\rm eff}=(N_{hs}-1)/N_{hs}\alpha'+B_{c}\delta_N/2N_{hs}$, which gives an interval for $10^{-2}\geq\xpom\gtrsim 10^{-4}$.}
    \label{tab:results}
\end{table}

\begin{table}[htbp]
\centering
\begin{subtable}[t]{0.35\textwidth}
\centering
\begin{tabular}{|l|rrrrr|}
\multicolumn{6}{c}{Fit 1: Gauss$\otimes K_0$ profile} \\\hline
 & $B_{c0}$ & $B_{hs}$ & $\lambda$ &
  $\sigma_{S}$ & $\alpha'$ \\
\hline
$B_{c0}$
    & 1.000 &       &        &        &       \\
$B_{hs}$
    & 0.405 & 1.000 &        &        &        \\
$\lambda$
    &-0.706 & -0.674 & 1.000  &        &       \\
$\sigma_{S}$
    &-0.734 &-0.183 & 0.751   & 1.000  &       \\
$\alpha'$
    &-0.049 & 0.002 & 0.010   & 0.038  & 1.000 \\
\hline
\end{tabular}
\end{subtable}
\begin{subtable}[t]{0.35\textwidth}
\centering
\begin{tabular}{|l|rrrrr|}
\multicolumn{6}{c}{Fit 2: Clem profile} \\
\hline
 & $B_{c0}$ & $\xi_v$ & $\lambda$ &
  $\sigma_{S}$ & $\alpha'$ \\
\hline
$B_{c0}$
    & 1.000 &       &        &        &       \\
$\xi_v$
    & 0.463 & 1.000 &        &        &        \\
$\lambda$
    &-0.693 &-0.800 & 1.000  &        &       \\
$\sigma_{S}$
    & -0.728 & -0.252 & 0.701 & 1.000  &       \\
$\alpha'$
    &-0.049 & 0.001 & 0.008  & 0.038  & 1.000 \\
\hline
\end{tabular}
\hfill
\end{subtable}
\caption{Correlation matrix of the fitted parameters in Fits 1 and 2 of table \ref{tab:results}}
\label{tab:correlation_matrix1}
\end{table}

In table \ref{tab:results}, Fits 1 and 2 show the fitted parameters for $N_{hs}=8$ with the squeezed proton. We report the parameter correlations for Fits 1 and 2 in table \ref{tab:correlation_matrix1}. 

We also show as Fit 3 the result from the additionally squeezed hotspots, with $B_{hs}=B_{hs0}+2\beta'\log\left({x_0}/{\xpom}\right)$. We see that the gain in overall fit quality is $\Delta\chi^2=1.25$ compared with Fit 1, which means that the HERA data cannot resolve this variation in hotspot geometry. We also see that the lower value of $\beta'=0.016(11)~{\rm GeV}^{-2}$ is nearly consistent with 0. In table \ref{tab:correlation_matrix2} we see that the $\beta'$ parameter is strongly correlated with $B_{hs}$ with a correlation of 0.941, and that $\alpha'$ has gained a correlation with the proton geometry parameters. 

\begin{table}[htbp]
\centering
\begin{subtable}[t]{0.35\textwidth}
\centering
\begin{tabular}{|l|rrrrrr|}
\multicolumn{7}{c}{Fit 3: $B_{hs}(\xpom)$} \\\hline
 & $B_{c0}$ & $B_{\mathrm{hs}}$ & $\lambda$ &
  $\sigma_{S}$ & $\alpha'$ & $\beta'$ \\
\hline
$B_{c0}$
    & 1.000 &       &        &        &        &       \\
$B_{\mathrm{hs}}$
    & 0.484 & 1.000 &        &        &        &       \\
$\lambda$
    &-0.734 &-0.800 & 1.000  &        &        &       \\
$\sigma_{S}$
    &-0.764 &-0.354 & 0.746  & 1.000  &        &       \\
$\alpha'$
    &-0.235 &-0.474 & 0.347  & 0.060  & 1.000  &       \\
$\beta'$
    & 0.388 & 0.941 &-0.676 & -0.314 & -0.504 & 1.000 \\
\hline
\end{tabular}
\end{subtable}
\qquad
\begin{subtable}[t]{0.35\textwidth}
\centering
\begin{tabular}{|l|rrrrrr|}
\multicolumn{7}{c}{Fit 4: $N_{hs}(\xpom)$} \\\hline
 & $B_{c0}$ & $B_{\mathrm{hs}}$ & $\lambda$ &
 $\sigma_{S}$ & $\alpha'$ & $\delta_N$ \\
\hline
$B_{c0}$
    & 1.000 &       &        &        &        &       \\
$B_{\mathrm{hs}}$
    & 0.418 & 1.000 &        &        &        &       \\
$\lambda$
    & -0.710 & -0.685 & 1.000  &        &        &       \\
$\sigma_{S}$
    &-0.663 & -0.301 & 0.701  & 1.000  &        &       \\
$\alpha'$
    &-0.118 &-0.191 & 0.120  & 0.420  & 1.000  &       \\
$\delta_N$
    & 0.105 & 0.271 &-0.155  &-0.563  &-0.697 & 1.000 \\
\hline
\end{tabular}
\end{subtable}
\caption{Correlation matrix of the fitted parameters of Fits 3 and 4 in table \ref{tab:results}.}
\label{tab:correlation_matrix2}
\end{table}

Following \cite{Cepila:2016uku}, we alternatively let the average number of hotspots vary with $\xpom$, as $\braket{N_{hs}}=N_{hs0}(x_0/\xpom)^{\delta_N}$, where the average $\braket{N_{hs}}$ replaces $N_{hs}$ in the cross section calculations. The results, shown in tab. \ref{tab:results} as Fit 4, are for $N_{hs0}=8$. Here, the gain in fit quality is negligible $\Delta\chi^2=0.1$, with $\delta_N=0.040(81)$, consistent with zero. This gives a hotspot variation of $\braket{N_{hs}}=7.35\dots8.84$ in the range $\xpom=10^{-2}\dots 10^{-4}$. The parameter correlations are reported in table \ref{tab:correlation_matrix2}. 

We see that the resulting $\alpha'_{\rm eff.}$ value is stable in all fits. 

\begin{figure}
    \centering
    \includegraphics[width=.98\linewidth]{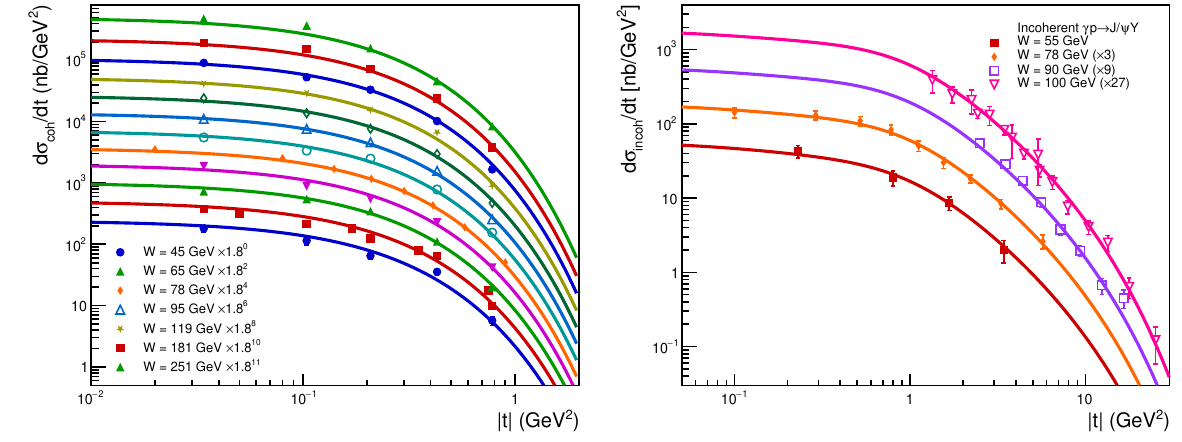}
    \caption{Comparison between Fit 1 with parameters from tab. \ref{tab:results}, and coherent and incoherent photoproduction H1 \cite{Aktas:2005xu,Alexa:2013xxa,Aktas:2003zi} and ZEUS \cite{Chekanov:2002rm,Chekanov:2009ab} measurements. The left side shows coherent data and the right side incoherent.}
    \label{fig:coherent}
\end{figure}

We show the results from Fit 1 in figure \ref{fig:coherent} for coherent and incoherent data. The description is good for the entire $t$-spectrum as well as for the $W$-spectrum. 

\subsection{Alternative Hotspot Shapes}
\begin{figure}
    \centering
    \includegraphics[width=.47\linewidth]{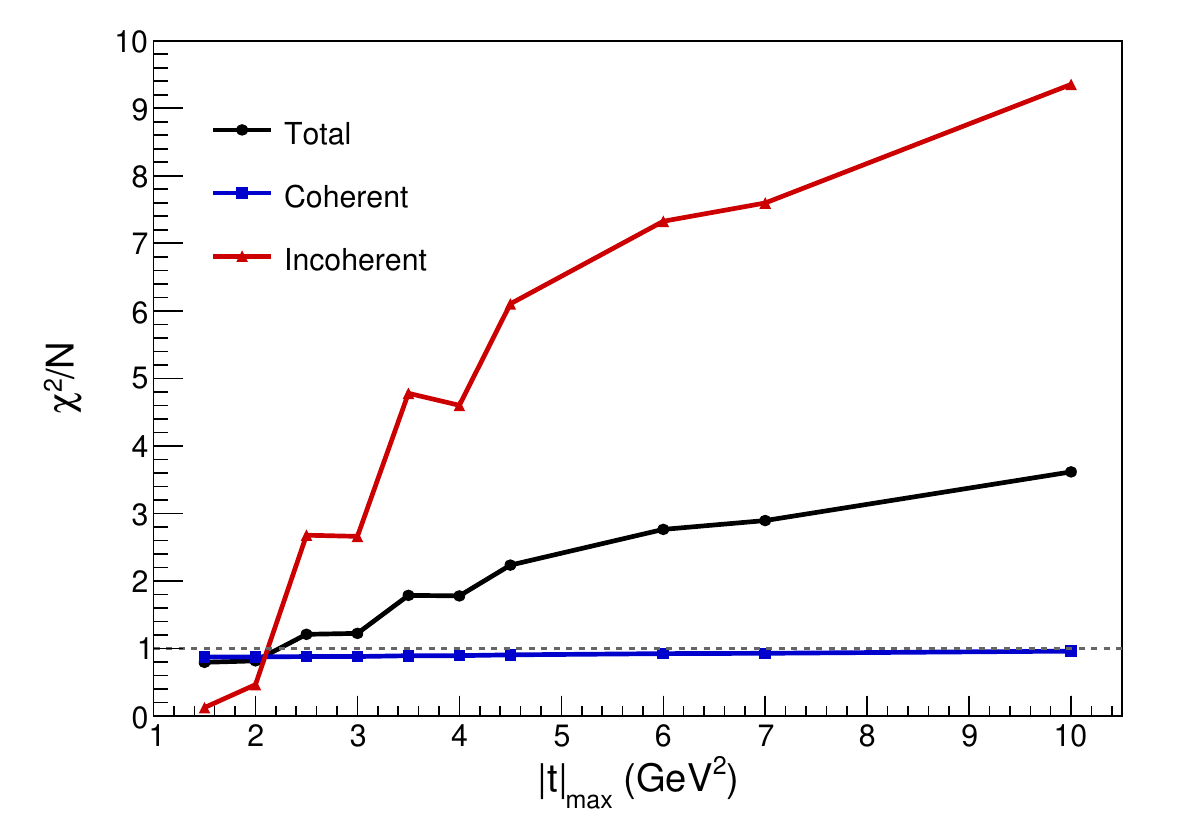}
    \caption{The resulting fit quality as a function of the maximum $|t|$ value included in the fit using a pure Gaussian hotspot profile, for the total quality as well as its coherent and incoherent components, with $N_{hs}=8$.}
    \label{fig:sigmaS}
\end{figure}
In previous studies, the hotspot shapes have been assumed to be Gaussian, and for larger values of $|t|$ they have been supplemented by substructure, such as point charges or smaller ``evolved''. hotspots. In fig.~\ref{fig:sigmaS} we show how the fit quality is affected by including more data up to $|t|_{\rm max}$. We see that the coherent cross section, which is completely contained in the region $|t|<1~{\rm GeV}^2$ is mostly unaffected by this, while for the incoherent cross section the description fails for $|t|\geq 2.5~{\rm GeV}^2$.

In order to ascertain that the hotspot shape, which is a convolution of a Gaussian with exponential tails, is what the data prefer, rather than it being the fact of having two parameters that control the hotspot geometry at different scales that allows for the description of the data, we also tried to fit some alternative shapes to the $t$-spectra. For this, we modified Fit 1 as described in section \ref{sec:profiles}.

We fitted the $n$ parameter in eq.~\eqref{eq:shape2}. The fit found $n=1.17(15)$ with a total $\chi^2=70.82$, or $\chi^2/{\rm ndf}=0.73$ (unchanged), which gives a total improvement of $\Delta\chi^2=0.94$. Therefore, $n$ is consistent with $n=1$ and the data does not provide significant evidence for $n\neq 1$. We also note that $n$ is strongly anti-correlated with $B_{hs}$ in the fit with a correlation of $-0.97$.
We also tried this fit without convoluting with the Gaussian core and found the best fit for $n=1.58(11)$, but the fit-quality is worsened by $\Delta\chi^2=5$, with the same number of parameters, which shows that the core belongs to the hotspot profile. 

We next fitted the stretched exponential formfactor in eq.~\eqref{eq:shape3}. Here, $\beta=0.49(4)$ which is consistent with $\beta=1/2$, corresponding to a power-law tail in the hotspot. However, the overall fit quality is significantly worse for this shape, with $\chi^2=101.69$, and $\chi^2/{\rm ndf}=1.05$, which gives a worsening in the fit quality of $\Delta \chi^2=29.9$. The corresponding power-law tails fall to the power of $3/2$. For $p\leq2$, the transverse second moment diverges, and the corresponding form factor does not have a finite derivative at $t=0$. Here, $\beta$ is strongly anti-correlated with $B_{hs}$ with a correlation of $-0.92$.

In order to test the power-law tails of the hotspot we directly fitted the two parameters $a$ and $p$ in eq.~\eqref{eq:shape4}. Note that in this shape there is no longer a $\lambda$ parameter so this fit has 7 parameters in total, which is the same as the previous two fits in this subsection. We found the best fit for $a=0.99(19)~{\rm GeV}^{-1}$ and $p=1.54(13)$, which means that the power-law prefers the same power as was found in the stretched exponential, $p=3/2$. Consequently, the fit quality did not improve, with $\chi^2=101.71$, and $\chi^2/{\rm ndf}=1.05$, the difference to Fit 1's quality is $\Delta\chi^2=30.0$. Both parameters $a$ and $p$ are strongly anti-correlated with $B_{hs}$ with correlations $-0.98$ and $-0.91$ respectively. 

For this study we conclude that the exponential tails are the preferred hotspot shape by the HERA data.

\section{Conclusions and Outlook}
\label{sec:conclusions}
We performed an analytic calculation of the coherent and incoherent cross section for exclusive $J/\psi$ production at HERA at leading twist. 
We take advantage of the fact that the calculation does not require numerical generation and averaging of thousands of hotspot configurations to describe the $t$-spectrum. We assumed that the hotspot centers are distributed according to a Gaussian, and that the hotspot shapes are a convolution between a Gaussian and a $K_0$ Bessel function, resulting in hotspots with hard Gaussian cores surrounded by exponential halos. We also fitted the Clem distribution to the same data, and found that the resulting hotspot profile is very similar to a Gaussian core with exponential tails. The Clem distribution has the benefit that it provides an interpretation of the fitted parameters as condensate penetration depth, and together with the core length the GL parameter $\kappa$ and the condensate correlation length can be determined.

We fitted the models to 104 H1 and ZEUS measurements (67 coherent and 37 incoherent) using six free parameters describing the proton geometry, hotspot profile, saturation-scale fluctuations, and overall normalization. We found an excellent fit, with a minimum for $N_{hs}=6\dots 7$ for the Gaussian core model for which $\chi^2/{\rm ndf}=71.66/98=0.73$, and a minimum at $N_{hs}=8\dots 9$ for the Clem model with $\chi^2/{\rm ndf}=70.82/98=0.72$.
Since the data is not sensitive to the number of hotspots in a large region, we settled for showing results for $N_{hs}=8$. The overlap between the best fit regions is $5\leq N_{hs}\leq 12$. 

The hotspot geometry remains stable throughout the fits, unchanging with $\xpom$, either as a Gaussian core with exponential tails or as the Clem distribution (which are similar enough to be considered different parametrizations of the same shape). It has a width of $B_{hs}=0.101(8)_{\rm fit}(12)_{N_{hs}}~{\rm GeV}^{-2}$ which translates into a hard core of size $\sqrt{2B_{hs}}=0.089(4)_{\rm fit}(6)_{N_{hs}}~{\rm fm}$, or $\xi_v=0.087(5)_{\rm fit}(11)_{N_{hs}}~{\rm fm}$, surrounded by an exponential halo of size $\lambda=0.217(5)_{\rm fit}(18)_{N_{hs}}~{\rm fm}$ for the Gaussian core, or $\lambda=0.191(5)_{\rm fit}(19)_{N_{hs}}$ for the Clem distribution. Both fits therefore agree within the $N_{hs}$ uncertainty and the average value is $\lambda=0.204(5)_{\rm fit}(18)_{N_{hs}}~{\rm fm}$. We predict that these results are independent of $Q^2$ which measurements at the future Electron-Ion Collider (EIC) or Large Hadron-electron Collider (LHeC)\cite{LHeC:2020van} will be able to confirm.

We calculated the proton's transverse gluon radius in both shapes and the result is consistent. The average value is $b_{\rm rms}=0.65(1)~{\rm fm}$ to be compared to the transverse charge radius of 0.6261(6) fm.

The extracted core scale $\xi_v$ clearly corresponds to the Gaussian core-size $B_{hs}$ with $\xi_v=\sqrt{2B_{hs}}$ for all $N_{hs}$ within uncertainties, while the extracted sizes of the soft halos are remarkably consistent and stable through all the fit in this paper. We get a Ginzburg-Landau parameter of $\kappa=2.7(3)_{\rm fit}(7)_{N_{hs}}$, which corresponds to the geometry of flux tubes in a Type-II dual superconductor. The penetration length $\lambda$ corresponds to a vector particle mass $m_V=1/\lambda=0.97(9)~{\rm GeV}$ (using the average $\lambda=0.20(2)~{\rm fm}$ from the fits) which is consistent with Lattice calculations of the gluelump mass $m_{1+-}=0.87(15)~{\rm GeV}$. We may note that in other hotspot models, such as \cite{Mantysaari:2022ffw, Le:2025bvq}, the authors introduce an infrared cut-off which acts as an exponential suppression, however, their value $m_{IR}\sim  0.506_{-0.356}^{+1.12}~{\rm GeV}$ is consistent with our $m_V$, though with a lower central value.  

Taking the extracted parameters from each best-fit region we get $\sqrt{2B_{hs}}/\xi_v=1.07(9)$. The core length corresponds to a mass $m_{\rm core}=1/\xi_v=2.3(3)~{\rm GeV}$, which is larger than the lightest glueball mass $m_{0++}=1.65(2)~{\rm GeV}$ by a factor 1.4(2). We can also use the Ginzburg-Landau framework to estimate the resulting string-tension for a dual superconductor, which gives the correct answer for $\alpha_S(1/\lambda^2)=0.36(2)$ in the LO $\overline{\rm MS}$ variable flavor scheme that we use in our models, to be compared to the matching value of $\alpha_S=0.37(9)$ indicating that the penetration depth sets the scale of the coupling in the flux tubes. 

We tested the gluon shape by convoluting the Gaussian core with three different tails. First we tried the form-factor $\propto (1+\lambda^2|t|)^{-n}$ and extracted $n=1.17(15)$ without improvement in the reduced $\chi^2$, and a strong anti-correlation between $n$ and $B_{hs}$ of -0.97, from which we conclude that $n=1$ is the preferred shape in this family of functions. Second we tried the stretched exponential form-factor and found the parameter $\beta=0.49(4)$ which is consistent with $\beta=1/2$, corresponding to power-law tails in the hotspot. However, the fit-quality was worsened with $\Delta\chi^2=30$ which means that the data has a clear preference for exponential tails to this distribution. Lastly we tried to directly fit a power-law tail and found the best fit for the power $p=1.54(13)$ which is consistent with $p=3/2$ corresponding to $\beta=1/2$ in the previous fit. Here, the model was equally disfavoured by the data with $\Delta\chi^2=30$.

We also fitted two energy dependence variations, where the hotspot geometry varies with $\xpom$, either the hotspot width or the average hotspot number. Neither variation yielded improved fits, from which we conclude that the HERA data is not able to resolve these variations. Even though these effects are not needed to describe HERA data, in a denser system such as at the EIC, or in UPC at the LHC, these effects may become important.

We get a pomeron $t$-slope $\alpha'_{\rm eff.}=0.060(19)_{\rm fit}(1)_{N_{hs}}~{\rm GeV}^{-2}$, from both fits, which is a lower value than what was found at the HERA experiments but consistent with BFKL results. This result is stable throughout all fits and is uncorrelated with the hotspot geometry. 

In the dual-superconductor picture of the QCD vacuum, the hotspots appear as a vacuum response to vacuum fluctuations of quarks and gluons acting as localised chromo-electric sources, where the core size $\xi_v$ specifies the scale at which the condensate is broken, corresponding to a scalar particle mass, while the penetration depth $\lambda$ is how deep the chromoelectric field penetrates into the vacuum, which corresponds to a massive vector particle (in regular superconductor, a massive photon). In a type-II superconductor, the vortex-lines repel at the scale of $\lambda$. This suggests a qualitative picture of the proton where two gluons at a distance $d<\xi$ in the transverse plane end up in the same hotspot, while gluons at a distance $\xi<d<\lambda$ end up in hotspots that repel to a distance $d>\lambda$. We found that the average gluon occupancy per hotspot is on the order of a few. This picture would suggest that the number of hotspots would increase with the gluon density, but HERA data could not see any such behaviour. We tested the prediction that hotspots repel by introducing a hotspot-hotspot pair repulsion and scanning its correlation length-scale. The data was not sensitive enough to either reject or confirm this addition. At the scale where such a repulsion would be seen in the incoherent spectrum $|t|\sim1/\lambda^2\sim 1~{\rm GeV}^2$, the spectrum is given largely by the saturation scale fluctuations with ${\rm Var}_{\rm sat.}/{\rm Var}_{\rm tot}\sim 0.6$. However, it would be possible for the EIC to measure this directly, by consistently tagging the proton dissociation mass $M_Y$, where large $M_Y$ events would be dominated by the saturation scale fluctuations, while small $M_Y$ events are dominated by the longer wave-length modes of the geometric fluctuations, and hence more sensitive to hotspot-hotspot correlations.

In future studies we will extend this study to higher twists which would enable us to study  heavy-ion targets as well as light vector meson production. In heavy ions, for example at the future EIC and the LHeC, or in current ultra-peripheral collisions at the LHC, the gluon densities are expected to be significantly larger, such that the gluon occupancy in each hotspot is large. We would expect this to show up in the incoherent $t$-spectrum, especially at large $|t|$ where the dipole would scatter directly with point sources. Then the physics would be potentially different from what is detailed in this paper.

This hotspot model can also serve as an input to flow calculation in heavy ion collisions at RHIC and the LHC.

\subsection*{Disclosure} AI tools, including Anthropic Claude Opus 5 and Sonnet 5, were used to assist with scientific reasoning, analytical derivations, code development and debugging, and manuscript preparation. The authors directed and independently verified all AI-assisted results, calculations, code, and scientific statements and are fully responsible for the content of the manuscript.

\subsection*{Acknowledgements}
We express our gratitude to Raju Venugopalan and Bj\"orn Schenke at BNL for insightful discussions. The authors acknowledge the support from the physics department of IIT Delhi. This research was supported by Core Research Grant (CRG) support CRG/2022/002507 from Anusandhan National Research Foundation (ANRF), Department of Science and Technology, Government of India.

\bibliographystyle{apsrev4-2}
\bibliography{Saturationandfluctuations}

\end{document}